\documentclass{ws-p8-50x6-00}

\begin{document}

\title{Multipole analysis for pion photoproduction with MAID and a dynamical
            models}

\author{ S. S. Kamalov$^1$, D. Drechsel$^2$, L. Tiator$^2$ and S. N. Yang$^3$}
\address{$^1$ Laboratory of Theoretical Physics, JINR, 141980
Dubna, Russia\\ $^2$Institut f\"ur Kernphysik, Universit\"at
Mainz, 55099 Mainz, Germany\\ $^3$Department of Physics, National
Taiwan University, Taipei, Taiwan}

\maketitle

\abstracts{ We present results of new analysis of pion
photoproduction data obtained with Dynamical and MAID models}

During the last few years we have developed and extended two
models for the analysis of pion photo and electroproduction, the
Dynamical Model~\cite{KY99} (hereafter called Dubna-Mainz-Taipei
(DMT) model) and the Unitary Isobar Model~\cite{MAID98} (hereafter
called MAID). The final aim of such an analysis is to shed more
light on the dynamics involved in nucleon resonance excitations
and to extract $N^*$ resonance properties in an unambiguous way.
For this purpose as a testing ground we will use benchmark data
bases recently created and distributed among different theoretical
groups.

The crucial point in a study of $N^*$ resonance properties is the
separation of the total amplitude (in partial channel
$\alpha=\{l,j\}$)
\begin{eqnarray}
t_{\gamma\pi}^{\alpha}=t_{\gamma\pi}^{B,\alpha}+t_{\gamma\pi}^{R,\alpha}
\end{eqnarray}
in background $t_{\gamma\pi}^{B,\alpha}$ and resonance
$t_{\gamma\pi}^{R,\alpha}$ contributions. In different theoretical
approaches this procedure is different, and consequently this
could lead to different treatment of the dynamics of $N^*$
resonance excitation. As an example we will consider the two
different models: DMT and MAID.

In accordance with Ref.\cite{KY99}, in the DMT model the
$t_{\gamma\pi}^{B,\alpha}$ amplitude is defined as
\begin{eqnarray}
t^{B,\alpha}_{\gamma\pi}(DMT)=e^{i\delta_{\alpha}}\cos{\delta_{\alpha}}
\left[v^{B,\alpha}_{\gamma\pi} + P\int_0^{\infty} dq'
\frac{q'^2\,R_{\pi
N}^{(\alpha)}(q_E,q')\,v^{B,\alpha}_{\gamma\pi}(q')}{E-E_{\pi
N}(q')}\right]\,,
\label{eq:backgr}
\end{eqnarray}
where $\delta_{\alpha}(q_E)$ and $R_{\pi N}^{(\alpha)}$ are the
$\pi N$ scattering phase shift and full $\pi N$ scattering
reaction matrix, in channel $\alpha$, respectively, $q_E$ is the
pion on-shell momentum. The pion photoproduction potential
$v^{B,\alpha}_{\gamma\pi}$ is constructed in the same way as in
Ref.\cite{MAID98} and contains contributions from the Born terms
with an energy dependent mixing of pseudovector-pseudoscalar
(PV-PS) $\pi NN$ coupling and t-channel vector meson exchanges. In
the DMT model $v^{B,\alpha}_{\gamma\pi}$ depends on 7 parameters:
The PV-PS mixing parameter $\Lambda_m$ (see Eq.(12) of
Ref.\cite{MAID98}), 4 coupling constants  and 2 cut-off parameters
for the vector mesons exchange contributions.

In the extended version of MAID, the $S$, $P$, $D$ and $F$ waves
of the background amplitudes $t_{\gamma\pi}^{B,\alpha}$ are
defined in accordance with the K-matrix approximation
\begin{equation}
 t_{\gamma\pi}^{B,\alpha}({\rm MAID})=
 \exp{(i\delta_{\alpha})}\,\cos{\delta_{\alpha}}
 v_{\gamma\pi}^{B,\alpha}(q,W,Q^2)\,,
\label{eq:bg00}
\end{equation}
where $W\equiv E$ is the total $\pi N$ c.m. energy and
$Q^2=-k^2>0$ is the square of the virtual photon 4-momentum. Note
that in actual calculations, in order to take account of inelastic
effects, the factor $\exp{(i\delta_{\alpha})}
\cos{\delta_{\alpha}}$ in Eqs.(2-3) is replaced by
$\frac{1}{2}[\eta_{\alpha}\exp{(2i\delta_{\alpha})} +1]$, where
both the $\pi N$ phase shifts $\delta_{\alpha}$ and inelasticity
parameters $\eta_{\alpha}$ are taken from the analysis of the SAID
group\cite{VPI97}.

From Eqs. (\ref{eq:backgr}) and (\ref{eq:bg00}), one finds that
the difference between the background terms of MAID and of the DMT
model is that pion off-shell rescattering contributions (principal
value integral) are not included in the background of MAID. From
our previous studies of the $P$ wave multipoles in the (3,3)
channel~\cite{KY99} it follows that they are effectively included
in the resonance sector leading to the dressing of the $\gamma
N\Delta$ vertex. However, in the case of $S$ waves the DMT results
show that off-shell rescattering contributions are very important
for the $E_{0+}$ multipole in the $\pi^0p$ channel. In this case
they have to be taken into account explicitly. Therefore, in the
extended version of MAID we have introduced a new phenomenological
term in order to improve the description of the $\pi^0$
photoproduction at low energies,
\begin{eqnarray}
E_{corr}(MAID)= \frac{\Delta E}{(1 + B^2q^2_{E})^2}\,F_D(Q^2)\,,
\label{eq:corr}
\end{eqnarray}
where $F_D$ is the standard nucleon dipole form factor, $B=0.71$
fm and $\Delta E$ is a free parameter which can be fixed by
fitting the low energy $\pi^0$ photoproduction data. Thus the
background contribution in MAID finally depends on 8 parameters.
Below $\pi^+n$ threshold for both models we also take into account
the cusp effect due to unitarity, as it was described in
Ref.\cite{Larget}, i.e.
\begin{eqnarray} E_{cusp}= - a_{\pi N} \,\omega_c\,Re
E_{0+}^{\gamma\pi^+}\,\sqrt{1-\frac{\omega^2}{\omega_c}}\,,
\label{eq:cusp}
\end{eqnarray}
where $\omega$ and $\omega_c=140$ MeV are the $\pi^+$ c.m.
energies corresponding to $W=E_p + E_{\gamma}$ and
$W_c=m_n+m_{\pi^+}$, respectively, and $a_{\pi N}=0.124/m_{\pi^+}$
is the pion charge exchange amplitude.

For the resonance contributions, following  Ref.\cite{MAID98}, in
both models the Breit-Wigner form is assumed, i.e.
\begin{equation}
t_{\gamma\pi}^{R,\alpha}(W,Q^2)\,=\,{\bar{\cal
A}}_{\alpha}^R(Q^2)\, \frac{f_{\gamma R}(W)\Gamma_R\,M_R\,f_{\pi
R}(W)}{M_R^2-W^2-iM_R\Gamma_R} \,e^{i\phi_R}\,,
\label{eq:BW}
\end{equation}
where $f_{\pi R}$ is the usual Breit-Wigner factor describing the
decay of a resonance $R$ with total width $\Gamma_{R}(W)$ and
physical mass $M_R$. The phase $\phi_R(W)$ in Eq. (\ref{eq:BW}) is
introduced to adjust the phase of the total multipole to equal the
corresponding $\pi N$ phase shift $\delta_{\alpha}$.

The main subject of our study in the resonance sector is the
determination of the strengths of the electromagnetic transitions
described by the amplitudes ${\bar{\cal A}}_{\alpha}^R(Q^2)$. In
general, they are considered as free parameters which have to be
extracted from the analysis of the experimental data. In our two
models we have included contributions from the 8 most important
resonances, listed in the Table 1. The total number of ${\bar{\cal
A}}_{\alpha}^R$ amplitudes is 12 and they can be expressed also in
terms of the 12 standard helicity elements $A_{1/2}$ and
$A_{3/2}$.
\begin{table}[htbp]
\begin{center}
\begin{tabular}{|cc|c|c|c|c|}
\hline $N^*$ &  &  MAID & MAID    &  DMT    & PDG2000   \\
      &  &  current    & HE fit & HE fit &  \\
\hline $P_{33}(1232)$ & $A_{1/2}$ & -138  & -143  & ---  &
-135$\pm$ 6 \\
               & $A_{3/2}$ & -256  & -264  & ---  & -255$\pm$ 8 \\
\hline $P_{11}(1440)$ & $A_{1/2}$ &  -71  & -81   &  -77  &
-65$\pm$ 4 \\ \hline $D_{13}(1520)$ & $A_{1/2}$ &  -17  & -6    &
-7   &  -24$\pm$ 9 \\
               & $A_{3/2}$ &  164  & 160  & 165  &  166$\pm$ 5 \\
\hline $S_{11}(1535)$ & $A_{1/2}$ &   67  &  81   &  102  &
90$\pm$ 30 \\ \hline $S_{31}(1620)$ & $A_{1/2}$ &   0  &  86    &
37   &  27$\pm$ 11 \\ \hline $S_{11}(1650)$ & $A_{1/2}$ &   39  &
32   &  34   & 53$\pm$ 16  \\ \hline $F_{15}(1680)$ & $A_{1/2}$ &
-10  &  5    &  10   & -15$\pm$ 6  \\
               & $A_{3/2}$ &  138  &  137  &  132  & 133 $\pm$ 12\\
\hline $D_{33}(1700)$ & $A_{1/2}$ &   86  & 119   &  107  &
104$\pm$ 15 \\
               & $A_{3/2}$ &   85  &  82   &  74   & 85$\pm$  22 \\
\hline
PV-PS mixing:   & $\Lambda_m$ &  450   &  406    & 302  &     \\
\hline
       &  $\Delta E$ &   2.01 & 1.73 & ---   &  \\
\hline
       &  $ \chi^2$/d.o.f. &   11.5 & 6.10 &  6.10  &     \\
\hline
\end{tabular}
\end{center}
\caption{Proton helicity amplitudes (in $10^{-3}\,GeV^{-1/2}$),
values of the PV-PS mixing parameter $\Lambda_m$ (in MeV) and
low-energy correction parameter $\Delta E$ (in
$10^{-3}/m_{\pi^+}$) obtained after the high-energy (HE) fit}
\end{table}
Thus, to analyze experimental data we have a total of 19
parameters in DMT and 20 parameters in MAID. The final results
obtained after the fitting of the high-energy (HE) benchmark data
base with 3270 data points in the photon energy range $180 <
E_{\gamma}<1200$ MeV are given in Table 1.

For the  analysis of the low-energy (LE) data base with 1287 data
points in the photon energy range  $180 < E_{\gamma}<450$ MeV in
the DMT model we used only 4 parameters: The PV-PS mixing
parameter and 3 parameters for the $P_{33}(1232)$ and
$P_{11}(1440)$ resonances. In MAID we have one more parameter due
to the low energy correction given by Eq. (\ref{eq:corr}). The
final results for the helicity elements and the E2/M1 ratio (REM)
are given in Table 2.
\begin{table}[htbp]
\begin{center}
\begin{tabular}{|cc|c|c|c|c|}
\hline
$N^*$ &  &  MAID & MAID    &  DMT    & PDG2000   \\
      &  &  current    & LE fit & LE fit &  \\
\hline
$P_{33}(1232)$ & $A_{1/2}$ & -138  & -142  & ---  & -135$\pm$ 6 \\
               & $A_{3/2}$ & -256  & -265  & ---  & -255$\pm$ 8 \\
\hline
$P_{11}(1440)$ & $A_{1/2}$ &  -71  & -81   &  -93  &  -65$\pm$ 4 \\
\hline
       &  REM(\%)   &   -2.2   &  -1.9    & -2.1  & -2.5$\pm$ 0.5\\
\hline
       &  $ \chi^2$/d.o.f. &   4.76 & 4.56 &  3.59  &  \\
\hline
\end{tabular}
\end{center}
\caption{Proton helicity elements (in $10^{-3}\,GeV^{-1/2}$) and
REM=E2/M1 ratio (in \%) obtained from the LE fit.}
\end{table}
In Table 3 we summarize our results and show the $\chi^2$ obtained
for different channels and different observables after fitting the
LE and HE data bases.
\begin{table}[htbp]
\begin{center}
\begin{tabular}{|c|ccc|ccc|}
\hline
     &       & LE     &         &        & HE   &     \\
Observables & N & MAID   &    DMT
&  N &  MAID   &    DMT    \\ \hline
$\frac{d\sigma}{d\Omega}(\gamma,\pi^+)$  &
 317 & 4.68    &   3.32  & 871 & 6.36  &   5.95  \\
$\frac{d\sigma}{d\Omega}(\gamma,\pi^0)$  &
 354 & 7.22    &   5.74  & 859 & 6.87  &   5.85  \\
\hline $\Sigma (\gamma,\pi^+)$  &
 245  & 2.79   &   2.57  &  546 & 4.57  & 6.49  \\
$\Sigma (\gamma,\pi^0)$  &
 192  & 2.22   &   1.58  & 488 &  7.65  & 7.65  \\
\hline $T (\gamma,\pi^+)$&
 107  & 3.28    &   2.94  & 265 & 3.75 &  4.17  \\
$T (\gamma,\pi^0)$  &
72 & 5.18    &   4.84  &  241 & 5.31  & 5.65  \\
\hline
  Total  & 1287 & 4.56    &   3.64  & 3270 & 6.10  &   6.10  \\
\hline
\end{tabular}
\end{center}
\caption{$\chi^2/N$ for the cross sections
($\frac{d\sigma}{d\Omega}$), photon ($\Sigma$) and target ($T$)
asymmetries in $(\gamma,\pi^+)$ and $(\gamma,\pi^0)$ channels
obtained after LE and HE fit. $N$ is the number of data points}
\end{table}
Note that the largest $\chi^2$ in the LE fit we get for
differential cross sections and target asymmetries in
$p(\gamma,\pi^0)p$. Similar results were obtained practically in
all other analyses. A detailed comparison with the results of
different theoretical groups is given on the website
$http://gwdac.phys.gwu.edu/analysis/pr\_benchmark.html$. Below, in
Fig. 1 we show only one interesting example, the $E_{0+}$
multipole in the channel with total isospin 1/2. In this channel
contributions from the $S_{11}(1535)$ and $S_{11}(1620)$
resonances are very important. At $E_{\gamma}>750$ MeV our values
for the real part of the $_pE_{0+}^{1/2}$ amplitude are mostly
negative and lower than the results of the SAID multipole
analysis. The only possibility to remove such a discrepancy in our
two models would be to introduce a third $S_{11}$ resonance.
Another interesting result is related to the imaginary part of the
$_pE_{0+}^{1/2}$ amplitude and, consequently, to the value of the
helicity elements given in Table 1. Within the DMT model for the
$S_{11}(1535)$ we obtain $A_{1/2}=102$ for a total width of 120
MeV, which is more consistent with the results obtained in $\eta$
photoproduction, than with previous pion photoproduction results
obtained by the SAID and MAID groups.
\begin{figure}[tbp]
\begin{center}
\epsfig{file=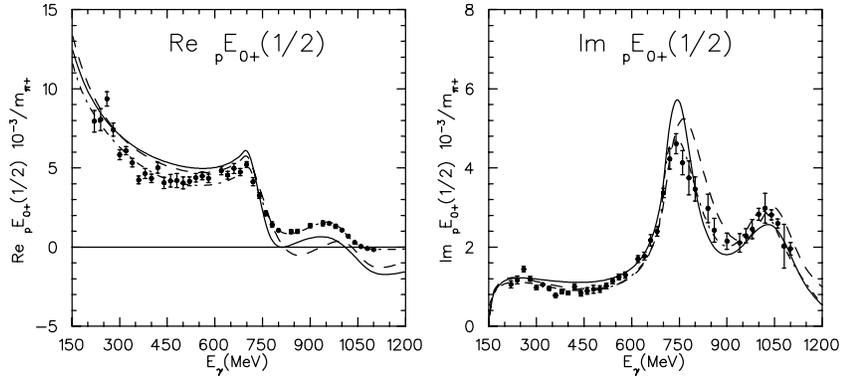, angle=90, width= 11 cm}
\end{center}
\caption{$_pE_{0+}^{1/2}$ multipole obtained after the HE fit
using MAID (solid curves) and DMT (dashed curves). The dash-dotted
curves and data points are the results of the global and
single-energy fits obtained by the SAID group.} \vspace*{-0.5cm}
%\caption{$_pE_{0+}^{1/2}$ multipole obtained after the HE fit
%using MAID (red curves) and DMT (blue curves). Green curves and
%data points are the results of global and single-energy fits
%obtained by the SAID group.} \vspace*{-0.5cm}
\end{figure}


\vspace*{-0.2cm}
\begin{thebibliography}{99}
\bibitem{KY99} S.S. Kamalov and S.N. Yang, \Journal{PRL}{83}{4494}{1999};
S.N. Yang, J. Phys. G {\bf 11}, L205 (1985).
\bibitem{MAID98} D. Drechsel, O. Hanstein, S.S. Kamalov and L. Tiator,
Nucl. Phys. {\bf A645}, 145 (1999).
\bibitem{VPI97} R.A. Arndt, I.I. Strakovsky and R.L. Workman,
 Phys. Rev. C {\bf 53}, 430 (1996).
\bibitem{Larget} J. M. Laget, Phys. Rep. {\bf 69}, 1 (1981).
\end{thebibliography}
\end{document}